\documentclass[aps,amssymb,amsmath,amsfonts,pra,twocolumn,superscriptaddress]{revtex4-1}
\usepackage[english]{babel}
\usepackage[latin1]{inputenc}
\usepackage{amssymb}
\usepackage{amsmath}
\usepackage[T1]{fontenc}
\usepackage{diagbox} 
\usepackage[usenames, dvipsnames]{color}
\usepackage{graphicx}
\usepackage{gensymb}
\usepackage{pifont}

\usepackage[caption=false]{subfig}
\usepackage[braket, qm]{qcircuit}
\usepackage{listings}
\usepackage{tasks}
\usepackage{hyperref}
\usepackage{mathtools}
\usepackage{cancel}
\newcommand{\overbar}[1]{\mkern 1.5mu\overline{\mkern-1.5mu#1\mkern-1.5mu}\mkern 1.5mu}
\newcommand{\floor}[1]{\left\lfloor #1 \right\rfloor}
\newcommand{\qwxo}[2][-1]{\ar @{-} [#1,0]_{#2}}

\DeclareGraphicsExtensions{.png,PNG,.pdf,.PDF}

\begin{document}
\title{Quantum circuits for maximally entangled states} 

\author{Alba Cervera-Lierta}
\email{a.cervera.lierta@gmail.com}
\affiliation{Barcelona Supercomputing Center (BSC).}
\affiliation{Dept. F\'{i}sica Qu\`{a}ntica i Astrof\'{i}sica, Universitat de Barcelona, Barcelona, Spain.}
\author{Jos\'{e} Ignacio Latorre}
\affiliation{Dept. F\'{i}sica Qu\`{a}ntica i Astrof\'{i}sica, Universitat de Barcelona, Barcelona, Spain.}
\affiliation{Nikhef Theory Group, Science Park 105, 1098 XG Amsterdam, The Netherlands.}
\affiliation{Center for Quantum Technologies, National University of Singapore, Singapore.}
\author{Dardo Goyeneche}
\affiliation{Dept. F\'{i}sica, Facultad de Ciencias B\'{a}sicas, Universidad de Antofagasta, Casilla 170, Antofagasta, Chile.}

\date{\today}

\begin{abstract}
We design a series of quantum circuits that generate absolute maximally entangled (AME) states to benchmark a quantum computer. A relation between graph states and AME states can be exploited to optimize the structure of the circuits and minimize their depth. Furthermore, we find that most of the provided circuits obey majorization relations for every partition of the system and every step of the algorithm. The main goal of the work consists in testing efficiency of quantum computers when requiring the  maximal amount of genuine multipartite entanglement allowed by quantum mechanics, which can be used to efficiently implement multipartite quantum protocols.
\end{abstract}
\maketitle


\section{Introduction}

There is a need to set up a thorough benchmarking strategy for quantum computers. Devices that operate in very different platforms are often characterized by the number of qubits they offer, their coherent time and the fidelities of one- and two-qubit gates. This is somewhat misleading as the performance of circuits are far below the expected when the amount of genuine multipartite entanglement contained in state is relatively high. 

There exist several figures of merit that try to quantify the success performance of a quantum device. Methods such as randomized benchmarking \cite{RB}, state and process tomography \cite{SPT} and gateset tomography \cite{GT1,GT2} are used to quantify gate fidelities. However, they are only useful for few-qubit experiments and fail when used to evaluate the performance of greater circuits \cite{fail1,fail2}.  In that sense, IBM proposed a metric to be used in arbitrary large quantum circuits called \textit{quantum volume} \cite{volume}. This figure takes into account several circuit variables like number of qubits, connectivity and gate fidelities. The core of the protocol is the construction of arbitrary circuits formed by one- and two-qubit gates that are complex enough to reproduce a generic $n$-qubit operation. One can expect to generate high entanglement in this kind of circuits. Even though, we should certify that this amount of entanglement will be large enough to perform some specific tasks that, precisely, demand high entanglement. A further relevant reference concerns to volumetric framework for quantum computer benchmarks \cite{BY19}.

Two reasonable ways to tests for quantum computers are the following: \emph{(i)} implement a protocol based on maximally entangled states, \emph{(ii)} solve a problem that is hard for a classical computer. These two ways are linked by the fact that quantum advantage requires large amounts of entanglement, so that classical computers are unable to carry the demanded task even when a sofisticated technique, e.g. tensor networks \cite{TN}, is considered. We believe that item \emph{(i)} is fully doable with the current state of the art of quantum computers, at least for a small number of qubits. On the other hand, item \emph{(ii)} is much more challenging, as classical computers efficiently work with a large number of bits.

Quantum correlations depend on the delicate balance of the coefficients of the wave function. It is natural to expect that quantum computers will have to be very refined to achieve such a good description of multipartite correlations along the successive action of gates. Entanglement is at the heart of quantum efficiency \cite{Jozsa}. Again, if a quantum computer is not able to generate faithful large entanglement, it will remain inefficient. 

A fundamental factor in quantum computing is the ability to generate large entangled states, such as area law violating states \cite{E13}. However, such ability has to be accomplished by a sufficiently large coherence time for such multipartite maximally entangled states. Note that  GHZ-like states are highly entangled and useful to test violate qubit Bell inequalities \cite{Mermin90}, but even more entangled are the Absolute Maximally Entangled (AME) states \cite{FFPP08,HCLR12,GALRZ15}, which are maximally entangled in every bipartition of the system. 

Following this line of thought, along this work we explore techniques to efficiently construct quantum circuits for genuinelly multipartite maximally entangled states. Some preliminar results reflect the difficulty to deal with quantum computers. For instance, the amount of Bell inequality violation rapidly decreases with the number of qubits considered \cite{Dani}.Also, the exact simulation of an analytical solvable model in a quantum computer significantly differs from the expected values, even when considering four qubits and less than thirty basic quantum gates \cite{Ising}. These examples illustrate the difference between gate fidelity and circuit fidelity, being the second one much harder to improve.  

We shall present quantum circuits required to build multipartite maximally entangled states. This proposal differs from bosonic sampling method, where large amounts of entanglement are faithfully reproduced by classical simulations \cite{Boson}. AME states find applications in multipartite teleportation \cite{HC13} and quantum secret sharing \cite{HCLR12,HC13}. Along our work, maximally entangled states will be exclusively used to test the strength of multipartite correlations in quantum computers. Another possible use, not discussed by us, could be generate quantum advantage with respect to a classical computer or, ideally, to achieve quantum supremacy.

We describe a benchmark suit of quantum circuits, where each one should deliver an AME state. The circuits provided were designed to minimize the number of required gates under the presence of restricted connectivity of qubits. Some of them consider individual systems composed by more than two internal levels each, which sometimes can be effectively reduced to qubits. In general, we consider simple and compact circuits, illustrating the way in which multipartite entanglement is generate step by step along the circuit. We have also pay attention to a simple criteria of majorization of the entropy of reductions, which basically implies that multipartite entanglement, quantified by the averaged entropy of reductions, monotonically increase for all partitions.

This work is organized as follows: In Section \ref{sec:Review}, we review the basic properties of AME states and show explicit examples. In Section \ref{sec:circuits}, we present the quantum circuits that generate AME states by using the properties of graph states. We also propose the simulation of AME states having local dimension larger than 2 by using qubits instead of qudits. In Section \ref{sec:major}, we analyze the entanglement majorization criteria in the proposed circuits and find further optimal circuits for experimental implementation, by imposing a majorization arrow in terms of entanglement. In Section \ref{sec:implementation}, we implement GHZ and AME states for five qubit systems in IBM quantum computers, quantify the state preparation quality by testing maximal violation of suitably chosen Bell inequalities. Finally, in Section \ref{sec:conclusions} we discuss and summarize the main results of the paper.

\section{Review of AME states}\label{sec:Review}

The study of AME states have become an intensive area of research along the last years due to both theoretical foundations and practical applications. In this section we briefly review the current state of the art of the field. For a more extensive review on AME states, see e.g. Ref. \cite{GALRZ15}.

\subsection{General properties of AME states}

AME states, also known in some references as \emph{maximally multipartite entangled states}, are $n$ qudit quantum states with local dimension $d$ such that every reduction to $\lfloor n/2\rfloor$ parties is maximally mixed, where $\lfloor\,\cdot\,\rfloor$ is the floor function. Such states are maximally entangled when considering the average entropy of reductions as a measure of multipartite entanglement. That is, when the average Von Neumann entropy $S(\rho)=\mathrm{Tr}[\rho\log\rho]$, taken over all reductions to $\lfloor n/2\rfloor$ parties, achieves the global maximum value $S(\rho)=\lfloor n/2\rfloor$, where logarithm is taken in basis $d$. For instance, Bell states and GHZ states are AME states for bipartite and three partite systems, respectively, for any number of internal levels $d$.

The existence of AME states for $n$ qudit systems composed by $d$ levels each, denoted AME($n,d$), is a hard open problem in general \cite{HESG18}. This problem is fully solved for any number of qubits: an AME($n,2$) exists only for $n=2,3,5,6$  \cite{S04,HGS17,AME42,AME72}.  Among all existing AME states, there is one special class composed by minimal support states. These states are defined as follows: an AME($n,d$) state has minimal support if it can be written as the superposition of $d^{\lfloor n/2\rfloor}$ fully separable orthogonal pure states. Here, we consider superposition at the level of vectors, in such a way that the linear combination of pure states always produce another pure state. For example, generalized Bell states for two-qudit systems and generalized GHZ states for three-qudit systems have minimal support. It is simple to show that all coefficients of every AME state having minimal support can be chosen to be identically equal to $d^{-\lfloor n/2\rfloor/2}$, i.e. identical positive numbers. By contrast, AME states having non-minimal support require to be composed by non-trivial phases in its entries in order to have all reduced density matrices proportional to the identity. In other words, non-minimal support AME states require destructive interference.

AME states connect to several mathematical tools. It is known that
AME states composed by $n$ parties and having minimal support, e.g. AME(2,2), AME(3,2) and AME(4,3), are one-to-one related to a special class of maximum distance separable (MDS) codes \cite{HP03}, index unity orthogonal arrays \cite{GZ14}, permutation multi-unitary matrices when $n$ is even \cite{GALRZ15} and to a set of $m=n-\lfloor n/2\rfloor$ mutually orthogonal Latin hypercubes of size $d$ defined in dimension $\lfloor n/2\rfloor$ \cite{GRDZ18}. 
On the other hand, AME states inequivalent to minimal support states, e.g. AME(5,2) or AME(6,2), are equivalent to quantum error correction codes \cite{S04}, quantum orthogonal arrays \cite{GRDZ18}, non-permutation multiunitary matrices \cite{GALRZ15} and $m=N-\lfloor n/2\rfloor$ mutually orthogonal quantum Latin hypercubes of size $d$ defined in dimension $\lfloor n/2\rfloor$ \cite{GRDZ18}.
 
AME states define an interesting mathematical problem itself but also they define attractive practical applications. These include quantum secret sharing \cite{HCLR12,HC13}, open destination quantum teleportation \cite{HC13} and quantum error correcting codes \cite{S04}, the last one being a fundamental ingredient for building a quantum computer.

\subsection{Explicit expressions of AME states}
The simplest AME($n,d$) state, denoted $\Omega_{n,d}$, having minimal support are the Bell and GHZ states 
\begin{eqnarray}
\Omega_{2,d}=\frac{1}{\sqrt{d}}\sum_{i=0}^{d-1}|ii\rangle,
\end{eqnarray}
and
\begin{eqnarray}
\Omega_{3,d}=\frac{1}{\sqrt{d}}\sum_{i=0}^{d-1}|iii\rangle,
\end{eqnarray}
respectively. These states are AME for any number of internal levels $d\geq2$. That is, every single particle reduction in both states $\Omega_{2,d}$ and $\Omega_{3,d}$, produces the maximally mixed state. On the other hand, it is not obvious to prove that there is no AME state for $n=4$ qubits \cite{AME42}. The AME(5,2) state \cite{LMPZ96} can be written as
\begin{equation}
|\Upsilon_{5,2}\rangle = \frac{1}{4\sqrt{2}} \sum_{i=1}^{32} c_i |i\rangle ,
\end{equation}
where the 5-digits binary decomposition of $i$ should be considered inside the ket and
\begin{align}
c_i=\{&1, 1, 1, 1, 1, -1, -1, 1, 1, -1, -1, 1, 1, 1, 1, 1, 1, 1, \nonumber\\
& -1, -1, 1, -1, 1, -1, -1, 1, -1, 1, -1, -1, 1, 1\}.
\end{align}
By using local unitary operations, the same state can be reduced to any of the following states \cite{HCLR12,Cleve99},
\begin{align}
 \ket{0_{L1}}= \frac{1}{4}
 (& \ket{00000}+\ket{10010}+\ket{01001}+\ket{10100} \nonumber\\
 &+ \ket{01010}-\ket{11011}-\ket{00110}-\ket{11000} \nonumber\\
 &- \ket{11101}-\ket{00011}-\ket{11110}-\ket{01111} \nonumber\\
 &- \ket{10001}-\ket{01100}-\ket{10111}+\ket{00101}),\nonumber\\
 \ket{1_{L1}}= \frac{1}{4}
 (& \ket{11111}+\ket{01101}+\ket{10110}+\ket{01011} \nonumber\\
 &+ \ket{10101}-\ket{00100}-\ket{11001}-\ket{00111} \nonumber\\
 &- \ket{00010}-\ket{11100}-\ket{00001}-\ket{10000} \nonumber\\
 &- \ket{01110}-\ket{10011}-\ket{01000}+\ket{11010}).
\end{align}

For $n=6$, an AME(6,2) state \cite{BPBZCP07} can be constructed from the above AME(5,2) states $|0_{L1}\rangle$ and $|1_{L1}\rangle$ as
\begin{align}
 \ket{\Omega_{6,2}}=& \frac{1}{\sqrt{2}}\left(\ket{0}\ket{0_{L1}}+\ket{1}\ket{1_{L1}}\right) \nonumber\\
=\frac{1}{4}\big(&\ket{000}(\ket{+-+}+\ket{-+-})- \nonumber\\
&\ket{001}(\ket{+--}-\ket{-++})+ \nonumber\\
&\ket{010}(\ket{++-}-\ket{--+})- \nonumber\\
&\ket{011}(\ket{+++}+\ket{---})- \nonumber\\
&\ket{100}(\ket{+++}-\ket{---})- \nonumber\\
&\ket{101}(\ket{++-}+\ket{--+})- \nonumber\\
& \ket{110}(\ket{+--}+\ket{-++})- \nonumber\\
&\ket{111}(\ket{+-+}-\ket{-+-})\big),
\end{align}
where $|\pm\rangle=\left(|0\rangle\pm|1\rangle\right)/\sqrt{2}$. This exemplifies that local unitaries can be used to find versions of an AME state with a reduced support. Similarly, an AME(5,2) state having eight real coefficients can be found by combining the two states 
\begin{align}
|0_{L2}\rangle &=\frac{1}{2}(|00000\rangle + |00011\rangle + |01100\rangle -|01111\rangle),\\
|1_{L2}\rangle &=\frac{1}{2}(|11010\rangle + |11001\rangle + |10110\rangle - |10101\rangle)
\end{align}
in the following way \cite{GALRZ15}:
\begin{equation}
|\Omega_{5,2}\rangle=\frac{1}{\sqrt{2}}\left(|0_{L2}\rangle+|1_{L2}\rangle\right).
\end{equation}
It can be shown that neither the five- nor six-qubit AME states have minimal support.

For systems composed by $n>3$ parties and $d>2$ internal levels it is not simple to construct AME states. The AME(4,3) state \cite{AMEGraph} is defined as follows
\begin{align}
  |\Omega_{4,3}\rangle &= \frac{1}{3}\sum_{i,j=0}^{2}|i\rangle|j\rangle|i+j\rangle|i+2j\rangle \nonumber\\
  &=\frac{1}{3}\big(| 0000 \rangle +| 0111 \rangle +| 0222 \rangle \nonumber\\
  &+| 1012 \rangle +| 1120  \rangle +| 1201 \rangle \nonumber\\
  &+ |2021 \rangle +| 2102 \rangle +|2210 \rangle  \big) .
\label{eq:AME43}
\end{align}  

In a similar way, we can derive the AME(6,4) state \cite{GZ14}:
\begin{align}
  \label{eq:AME64}
   |\Omega_{6,4}\rangle =\frac{1}{8} & \big( \ket{000000}+ \ket{111100}+ \ket{222200}+ \ket{333300}+ \nonumber\\
   &\ket{321010}+ \ket{230110}+ \ket{103210}+ \ket{012310}+ \nonumber \\ 
&\ket{132020}+ \ket{023120}+ \ket{310220}+ \ket{201320}+ \nonumber\\
& \ket{213030}+ \ket{302130}+ \ket{031230}+ \ket{120330}+ \nonumber \\ 
&\ket{231001}+ \ket{320101}+ \ket{013201}+ \ket{102301}+ \nonumber\\
&\ket{110011}+ \ket{001111}+ \ket{332211}+ \ket{223311}+ \nonumber\\
&\ket{303021}+ \ket{212121}+ \ket{121221}+ \ket{030321}+ \nonumber\\
& \ket{022031}+ \ket{133131}+ \ket{200231}+ \ket{311331}+ \nonumber\\
&\ket{312002}+ \ket{203102}+ \ket{130202}+ \ket{021302}+ \nonumber\\
&\ket{033012}+ \ket{122112}+ \ket{211212}+ \ket{300312}+ \nonumber \\ 
&\ket{220022}+ \ket{331122}+ \ket{002222}+ \ket{113322}+ \nonumber\\
& \ket{101032}+ \ket{010132}+ \ket{323232}+ \ket{232332}+ \nonumber \\ 
&\ket{123003}+ \ket{032103}+ \ket{301203}+ \ket{210303}+ \nonumber\\
&\ket{202013}+ \ket{313113}+ \ket{020213}+ \ket{131313}+ \nonumber \\ 
&\ket{011023}+ \ket{100123}+ \ket{233223}+ \ket{322323}+ \nonumber\\
&\ket{330033}+ \ket{221133}+ \ket{112233}+ \ket{003333}\big) \, .
\end{align} 
This state is formed by $4^3=64$ equally superposed orthogonal states, so it is an AME state of minimal support.

\section{Quantum Circuits to construct AME states}\label{sec:circuits}

As mentioned above, AME states can be constructed in different ways. For our purpose, we consider graph states formalism \cite{HEB04}. Graph states are represented by an undirected graph, where each vertex corresponds to a $|+\rangle$ state 
and each edge with a Control-Z (CZ) gate. We can easily construct the quantum circuit for a graph state by considering a simple rule, as we will see later. In addition, a graph can be transformed into another -equivalent one- by applying local unitary operations \cite{LC}. This kind of transformation modifies the number of edges of a graph but not its entanglement properties. This property could allows us to adapt the circuit to different quantum chip architectures, in order to reduce as much as possible the number of gates required to physically implement the state.

Despite graph states can be defined in any local dimension $d$, quantum computers can only implement qubit quantum circuits. Nonetheless, we can simulate AME states having larger local dimensions $d$ by using qubits. That is, by mapping each qudit state into a multi-qubit state and by adapting $d$-dimensional gates into non-local qubit gates, as we explain in Subsection \ref{subsec:AMEfromcircuits}.

\subsection{Graph States}

Graph states are $n$ partite pure quantum states constructed from an undirected graph composed by $n$ vertices $\mathcal{V}=\{v_{i}\}$ and connected by edges $\mathcal{E}=\{e_{ij}=\{v_{i},v_{j}\}\}$. Each graph has associated an adjancency matrix $A$, whose entries satisfy that $A_{ij}=1$ if an edge $e_{ij}$ exists and $A_{ij}=0$ otherwise. Self-interactions are forbidden, meaning that diagonal entries of $A$ vanish.

A graph state for $n$ qudits can be constructed as follows \cite{HEB04,AMEGraph}:
\begin{equation}\label{graphstate}
|G\rangle=\prod_{i<j}^{n}\mathrm{CZ}_{ij}^{A_{ij}}(F_{d}|\bar{0}\rangle)^{\otimes{n}},
\end{equation} 
where
\begin{equation}
\mathrm{CZ}_{ij}=\sum_{k=0}^{d-1}\omega^{kl}|\bar{k}\rangle\langle \bar{k}|_{i}\otimes|\bar{l}\rangle\langle \bar{l}|_{j},
\label{eq:CZgate}
\end{equation}
is the generalized controlled-$Z$ gate, $\omega=e^{2\pi i/d}$ and
\begin{equation}
F_{d}=\frac{1}{\sqrt{d}}\sum_{k=0}^{d-1}\omega^{kl}|\bar{k}\rangle\langle \bar{l}|,
\label{eq:Fgate}
\end{equation}
is the Fourier qudit gate. From now on, we distinguish between qubits and qudits states, by writing a bar over symbols associated to qudit states, e.g. $|\bar{0}\rangle$, keeping the usual notation with no bar for qubits, e.g. $|0\rangle$.

\begin{figure}[t!]
\subfloat[]{\label{Fig:AME52a}
$
\raisebox{9em}{\Qcircuit @C=1.1em @R=.7em 
{
\lstick{\ket{0}} & \gate{H} & \ctrl{1} & \qw & \qw & \qw & \ctrl{4} & \qw \\
\lstick{\ket{0}} & \gate{H} & \ctrl{-1} & \ctrl{1} & \qw & \qw & \qw & \qw \\
\lstick{\ket{0}} & \gate{H} & \qw & \ctrl{-1} & \ctrl{1} & \qw & \qw & \qw \\
\lstick{\ket{0}} & \gate{H} & \qw & \qw & \ctrl{-1} & \ctrl{1} & \qw & \qw \\
\lstick{\ket{0}} & \gate{H} & \qw & \qw & \qw & \ctrl{-1} & \ctrl{-4}  & \qw
}}
$}\hspace{0.5cm}
\subfloat[]{
\includegraphics[width=0.4\columnwidth]{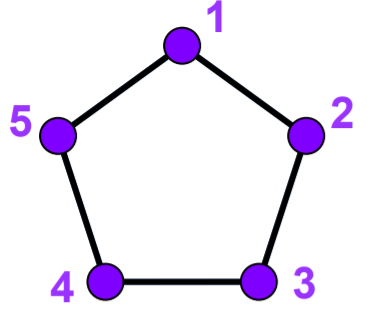}
\label{Fig:AME52b}
}
\caption{Quantum circuit to generate AME(5,2) (a) and its corresponding graph (b).}
\label{Fig:AME52}
\end{figure}

Following the above definition, the explicit construction of a graph state from its corresponding graph is simple. First, each vertex corresponds with the qudit state $|\bar{\psi}_{0}\rangle = F_{d}|\bar{0}\rangle$, and second, each edge corresponds with a CZ gate applied between two vertices. For instance, consider the quantum circuit generating the AME(5,2) state, see Figure \ref{Fig:AME52}. For qubits, note that $F_{2}$ gate is actually the Hadamard gate. Preparation of a qubit graph state (\ref{graphstate}) is equivalent to initialize all qubits in the state $|+\rangle=(|0\rangle+|1\rangle)/\sqrt{2}$ and then apply CZ gates between the qubits, according to the chosen graph.

Note that after applying the Fourier gate $F_d$ over the initial state $|\bar{0}\rangle^{\otimes n}$ we obtain a state with all basis elements, in the computational basis decomposition. Then, since the CZ gates only introduce relative phases between these elements, the final state of a graph contains a superposition of the $d^n$ elements of the computational basis. 

Graph states can also be described by using stabilizer states \cite{stab}. They find application in quantum error correcting codes \cite{QEC} and one-way quantum computing \cite{oneway}. A graphical interpretation of entanglement in graph states is provided in Ref.\cite{AMEGraph} and multipartite entanglement properties in qubit graph states, as well as its optimal state preparation, has been studied in Ref. \cite{HEB04,Cabello,OptimalGraph}.

\subsection{AME states from graph states}

We can write an AME state by using its corresponding graph, as described above. This is a particular form of an AME state having maximal support, as we have the superposition of all elements of the computational basis. 

We are interested in finding optimal AME graph states, in the sense of having the minimum number of edges and coloring index \cite{Cabello}. The smaller the number of edges the smaller the number of operations required to generate AME states. Coloring index is related with the number of operations that can be performed in parallel, so it is proportional to the circuit depth. It worth to mention that graph AME states are hard to construct in general, specially for large values of local dimension $d$ and number of parties $n$. Fortunately, there are suitable tools useful to simplify the construction of graphs for specific values of $d$ and $n$ \cite{AMEGraph}.

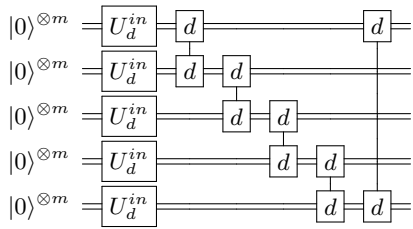
\begin{figure}[t!]
\centering
$
\Qcircuit @C=0.8em @R=.1em 
{
 \lstick{\ket{0}^{\otimes m}} & \gated{U_d^{in}}& \gated{d}  & \cw & \cw & \cw & \gated{d} & \cw \\
 & & \qwxo{\hspace{1pt}} & & & &\qwxo{\hspace{1pt}} & \\
 \lstick{\ket{0}^{\otimes m}} & \gated{U_d^{in}}&\gated{d} \qwxo{\hspace{1pt}} & \gated{d} & \cw & \cw & \qwxo{\hspace{1pt}} \cw & \cw \\
  &  & & \qwxo{\hspace{1pt}} & & & \qwxo{\hspace{1pt}} & \\
  \lstick{\ket{0}^{\otimes m}} & \gated{U_d^{in}}& \cw & \gated{d} \qwxo{\hspace{1pt}} & \gated{d} & \cw & \cw \qwxo{\hspace{1pt}} & \cw\\
 &  & & &  \qwxo{\hspace{1pt}} & & \qwxo{\hspace{1pt}} & \\
 \lstick{\ket{0}^{\otimes m}} & \gated{U_d^{in}}& \cw  & \cw & \gated{d} \qwxo{\hspace{1pt}} & \gated{d} & \cw \qwxo{\hspace{1pt}} & \cw\\
  &  & &  & & \qwxo{\hspace{1pt}} & \qwxo{\hspace{1pt}} & \\
  \lstick{\ket{0}^{\otimes m}} & \gated{U_d^{in}}& \cw  & \cw & \cw & \gated{d} \qwxo{\hspace{1pt}} & \gated{d} \qwxo{\hspace{1pt}}& \cw 
}$
\caption{Quantum circuit to generate an AME(5,d) state by using qubits instead of qudits. The corresponding graph is the same as the one in Fig. \ref{Fig:AME52b}. The number of qubits needed to represent each qudit is $m=\lceil\log_{2}d\rceil$. First, qubits are prepared in the basis superposition state by using $U_{d}^{in}$, which corresponds to $H$ for qubits, $U_{3}^{in}$ of Fig. \ref{Fig:F3} for qutrits and $U_{4}^{in}=H\otimes H$ for ququarts. Then, CZ gates are performed between the qudits, which for $d=3$ and $d=4$ can be implemented with the circuit shown in Fig. \ref{Fig:CZ3} and \ref{Fig:CZ4}, respectively.}
\label{Fig:AME5d}
\end{figure}

\begin{figure}[t!]
\subfloat[]{
$\hspace{1cm}
\raisebox{6.5em}{\Qcircuit @C=1em @R=.1em 
{
 \lstick{\ket{0}^{\otimes m}} & \gated{U_d^{in}}& \gated{d} & \cw & \cw & \cw & \gated{d} & \cw    \\
 & & \qwxo{\hspace{1pt}} & & & & \qwxo{\hspace{1pt}} &  \\
 \lstick{\ket{0}^{\otimes m}} & \gated{U_d^{in}}&\gated{d} \qwxo{\hspace{1pt}} & \gated{d} & \gated{d} & \cw & \cw \qwxo{\hspace{1pt}} & \cw  \\
  &  &  & \qwxo{\hspace{1pt}} & \qwxo{\hspace{1pt}} & & \qwxo{\hspace{1pt}} & \\
  \lstick{\ket{0}^{\otimes m}} & \gated{U_d^{in}}& \cw & \cw \qwxo{\hspace{1pt}} &   \cw\qwxo{\hspace{1pt}} & \gated{d} & \gated{d} \qwxo{\hspace{1pt}} & \cw \\
 &  &  & \qwxo{\hspace{1pt}} & \qwxo{\hspace{1pt}} & \qwxo{\hspace{1pt}} & & \\
 \lstick{\ket{0}^{\otimes m}} & \gated{U_d^{in}}& \cw  & \gated{d}\qwxo{\hspace{1pt}} & \gated{d}\qwxo{\hspace{1pt}} & \gated{d}\qwxo{\hspace{1pt}} & \cw & \cw
}}
$
\label{Fig:AME4da}}
\subfloat[]{
\includegraphics[width=0.13\textwidth]{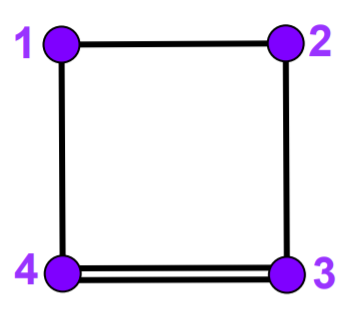}
\label{Fig:AME4db}}
\caption{Graph state that generates an AME(4,$d$) state for any prime dimension $d\geq 3$ (b) and its corresponding circuit (a) by using qubits instead of qudits.}
\label{Fig:AME4d}
\end{figure}

\begin{figure}[t!]
\centering
\subfloat[]{
$\hspace{1cm}
\Qcircuit @C=0.8em @R=.1em 
{
 \lstick{\ket{0}^{\otimes m}} & \gated{U_d^{in}}& \gated{d}  & \cw & \cw & \cw & \cw & \gated{d} & \gated{d} & \cw & \cw & \cw \\
 & & \qwxo{\hspace{1pt}} & & & & & \qwxo{\hspace{1pt}} & \qwxo{\hspace{1pt}} & & & & \\
 \lstick{\ket{0}^{\otimes m}} & \gated{U_d^{in}}& \gated{d} \qwxo{\hspace{1pt}} & \gated{d} & \cw & \cw & \cw & \qwxo{\hspace{1pt}} \cw & \cw \qwxo{\hspace{1pt}} & \gated{d} & \cw & \cw \\
  &  & & \qwxo{\hspace{1pt}} & & & & \qwxo{\hspace{1pt}} & \qwxo{\hspace{1pt}} & \qwxo{\hspace{1pt}} & \\
  \lstick{\ket{0}^{\otimes m}} & \gated{U_d^{in}}& \cw & \gated{d} \qwxo{\hspace{1pt}} & \gated{d} & \cw & \cw & \cw \qwxo{\hspace{1pt}} & \cw \qwxo{\hspace{1pt}} & \cw \qwxo{\hspace{1pt}} & \gated{d} & \cw \\
 &  & & &  \qwxo{\hspace{1pt}} & & & \qwxo{\hspace{1pt}} & \qwxo{\hspace{1pt}} & \qwxo{\hspace{1pt}} & \qwxo{\hspace{1pt}} & \\
 \lstick{\ket{0}^{\otimes m}} & \gated{U_d^{in}}& \cw  & \cw & \gated{d} \qwxo{\hspace{1pt}} & \gated{d} & \cw & \cw \qwxo{\hspace{1pt}} & \cw \qwxo{\hspace{1pt}} & \gated{d} \qwxo{\hspace{1pt}} & \cw \qwxo{\hspace{1pt}} & \cw\\
  &  & &  & & \qwxo{\hspace{1pt}} & & \qwxo{\hspace{1pt}} & \qwxo{\hspace{1pt}} & & \qwxo{\hspace{1pt}} & & \\
  \lstick{\ket{0}^{\otimes m}} & \gated{U_d^{in}}& \cw  & \cw & \cw & \gated{d} \qwxo{\hspace{1pt}} & \gated{d} & \cw \qwxo{\hspace{1pt}}& \gated{d} \qwxo{\hspace{1pt}} & \cw & \cw \qwxo{\hspace{1pt}} & \cw \\
   &  & &  & &  & \qwxo{\hspace{1pt}} & \qwxo{\hspace{1pt}} & & & \qwxo{\hspace{1pt}}\\
  \lstick{\ket{0}^{\otimes m}} & \gated{U_d^{in}}& \cw  & \cw & \cw & \cw & \gated{d} \qwxo{\hspace{1pt}} & \gated{d} \qwxo{\hspace{1pt}} & \cw & \cw & \gated{d} \qwxo{\hspace{1pt}} & \cw
}
$
\label{Fig:AME6da}}\\
\subfloat[]{
\includegraphics[width=0.4\columnwidth]{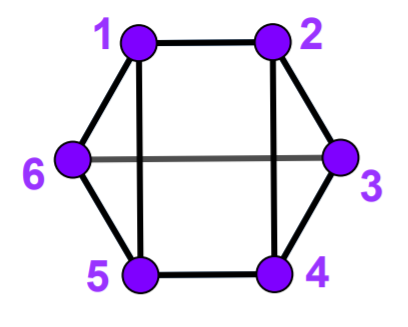}
\label{Fig:AME6db}}
\caption{Graph state that generates an AME(6,$d$) state (b) and its corresponding circuit (a) by using qubits instead of qudits. The number of qubits needed for represent each qudit is $m=\lceil\log_{2}d\rceil$. Qubits are prepared by using $U_{d}^{in}$ and CZ gates in dimension $d$, simulated by using the circuits shown in Fig. \ref{Fig:CZ3} and \ref{Fig:CZ4}.}
\label{Fig:AME6d}
\end{figure}

The first interesting property is that some graph states can be constructed in any dimension $d$. The simpler cases are given by the generalized Bell ($n=2$) and generalized GHZ states ($n=3$). The graph states of $n=5$ and $n=6$, shown in Figures \ref{Fig:AME5d} and \ref{Fig:AME6d} respectively, produce AME states in any prime dimension $d$. The $n=4$ graph state of Figure \ref{Fig:AME4d} also fulfills this property for every prime dimension $d\geq 3$.

\begin{figure}[t!]
\subfloat[]{
$
\hspace{0.6cm}
\raisebox{10em}{\Qcircuit @C=1em @R=.1em 
{
\lstick{\ket{0}} & \gate{H} & \qw & \qw & \qw & \qw & \ctrl{3} & \ctrl{4} & \qw & \qw & \qw  \\
\lstick{\ket{0}} & \gate{H} & \ctrl{1} & \qw & \qw & \qw & \qw & \qw & \qw & \ctrl{5} & \qw  \\
\lstick{\ket{0}} & \gate{H} & \ctrl{-1} & \ctrl{3} & \qw & \qw & \qw & \qw & \qw & \qw & \qw  \\
\lstick{\ket{0}} & \gate{H} & \qw & \qw & \qw & \ctrl{4} & \ctrl{-3} & \qw & \qw & \qw & \qw  \\
\lstick{\ket{0}} & \gate{H} & \qw & \qw & \qw & \qw & \qw & \ctrl{-4} & \ctrl{2} & \qw & \qw  \\
\lstick{\ket{0}} & \gate{H} & \qw & \ctrl{-3} & \ctrl{2} & \qw & \qw & \qw & \qw & \qw & \qw   \\
\lstick{\ket{0}} & \gate{H} & \qw & \qw & \qw & \qw & \qw & \qw & \ctrl{-2} & \ctrl{-5} & \qw  \\
\lstick{\ket{0}} & \gate{H} & \qw & \qw & \ctrl{-2} & \ctrl{-4} & \qw & \qw & \qw & \qw & \qw 
}}
$
\label{Fig:AME44_circuita}}
\subfloat[]{
\includegraphics[width=0.18\textwidth]{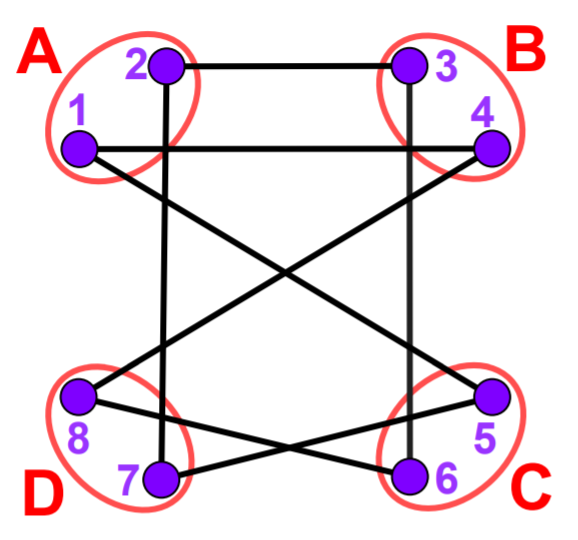}
\label{Fig:AME44_circuitb}}
\caption{Quantum circuit producing the AME(4,4) state with qubits (a) and its corresponding graph (b). Parties $A$, $B$, $C$ and $D$ are maximally entangled between them but not the qubits inside each party.
Notice that this circuit does not correspond to an AME(8,2) state, since this AME state does not exist.}
\label{Fig:AME44_circuit}
\end{figure}

For a non-prime local dimension there exist some methods to find AME graph states \cite{AMEGraph}. One of those consists on taking the prime factorization $d=d_{1}d_{2}\cdots d_{m}$ and look for every AME($n,d_{i}$) state. The AME($n,d$) is just given by the tensor product of the AME($n,d_{i}$) states, followed by a suitable relabeling of symbols. When prime factorization of $d$ includes a power of some factor, we can construct an AME state by artificially defining each party, i.e. by using qudits in lower dimension $m<d$ and then performing the suitable set of CZ gates between the $m$ level qudit systems. For instance, this method can be used to find an the AME(4,4) state from qubits instead of ququarts (qudits with $d=4$ levels each), as we illustrate in Figure \ref{Fig:AME44_circuit}. The -real- local dimension of each party, $d=4$, is achieved by grouping qubits in pairs \cite{AMEGraph}.

\subsection{AME states circuits using qubits}\label{subsec:AMEfromcircuits}

As we have seen above, Bell and GHZ states together with the graphs from Figures \ref{Fig:AME5d} to \ref{Fig:AME44_circuit} serve to construct AME($n,d$), for $n=2-6$ and   prime number of internal levels $d\geq 3$, and also the AME(4,4) state. Moreover, we can use a combination of these graphs to construct AME states of greater levels $d$.

The construction of a qubit quantum circuit from a graph state is straightforward since we just have to perform Hadamard gates on all qubits initialized at $|0\rangle$ state and CZ gates, according to graph edges. These quantum gates are commonly used in current quantum devices, e.g. in quantum computing \cite{IBM}. However, in order to implement an AME state for $d>2$ internal levels we require a qudit quantum computer, i.e. a machine performing quantum operations beyond binary quantum computation. The construction of such device is much more challenging than the current quantum computers and, therefore, perform such kind of experiment become really hard. Here, we propose to simulate AME states having larger local dimension by using qubits instead of qudits. To do so, we translate the local dimension $d$ into a multiqubit dimensional space. For instance, to transform a ququart system $d=4$ into a two qubit system $m=2$ we consider the following identification
\begin{equation}
|\bar{0}\rangle\equiv |00\rangle, \quad |\bar{1}\rangle\equiv |01\rangle,  \quad |\bar{2}\rangle\equiv |10\rangle,  \quad|\bar{3}\rangle\equiv |11\rangle.
\label{eq:ABC}
\end{equation}
For $d>4$, we need to increase the number of qubits accordingly, i.e. we need $m=\lceil\log_{2} d\rceil$ qubits to describe each $d$-level system, where $\lceil\,\cdot\,\rceil$ denotes the ceiling function. 
Since we have the graphs for these states, the challenge is to simulate the effect of the generalized CZ gate \eqref{eq:CZgate} and the Fourier gate \eqref{eq:Fgate}, with qubit gates. To be precise, we are not interested in the exact Fourier gate but on generating the state $|\bar{\psi}_{0}\rangle = F_{d}|\bar{0}\rangle$. For that propose, we will look for an initialization gate $U_{d}^{in}$ that acts on qubits in the state $|0\rangle$ and obtains the $|\psi_{0}\rangle$ state, i.e. the state $|\bar{\psi}_{0}\rangle$ written in terms of qubits according to the mapping established by Eq.\eqref{eq:ABC}.

When local dimension $d$ is a power of 2, the state $|\bar{\psi}_{0}\rangle$ can be easily generated by using Hadamard gates only. In particular, for $d=4$ we have
\begin{align}
|\bar{\psi}_{0}\rangle &=F_{4}|\bar{0}\rangle=\frac{1}{2}\left(|\bar{0}\rangle+|\bar{1}\rangle+|\bar{2}\rangle+|\bar{3}\rangle\right)\nonumber \\
|\psi_{0}\rangle & = U_{4}^{in}|00\rangle = (H\otimes H)|00\rangle \nonumber\\ &=\frac{1}{2}\left(|00\rangle+|01\rangle+|10\rangle+|11\rangle\right).
\end{align}
Despite $F_{4}\neq U_{4}^{in}=(H\otimes H)$, the tensor product unitary transformation is suitable, as we just want to obtain the state $|\bar{\psi}_{0}\rangle$ with qubits.

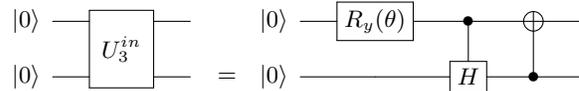
\begin{figure}[t!]
\[
\Qcircuit @C=1.5em @R=.7em @!R {
 \lstick{\ket{0}} & \multigate{1}{U^{in}_{3}} & \qw & & & \lstick{\ket{0}} & \gate{R_y(\theta)} &  \ctrl{1} & \targ & \qw \\
\lstick{\ket{0}} & \ghost{U^{in}_{3}} & \qw & = &
 & \lstick{\ket{0}} & \qw & \gate{H} & \ctrl{-1} & \qw
}
\]
\caption{Quantum circuit to obtain $|\bar{\psi}_{0}\rangle$ qutrit state using two qubits, i.e. to generate $|\psi_{0}\rangle=\left(|00\rangle+|01\rangle+|10\rangle\right)/\sqrt{3}$ state. The angle of the rotational gate is $\theta=-2\arccos(1/\sqrt{3})$.}
\label{Fig:F3}
\end{figure}

For $d=3$ the state $|\bar{\psi}_{0}\rangle$ can be obtained from the gate $U_{3}^{in}$, defined in Figure \ref{Fig:F3}:
\begin{align}
|\bar{\psi}_{0}\rangle &= F_{3}|\bar{0}\rangle=\frac{1}{\sqrt{3}}\left(|\bar{0}\rangle+|\bar{1}\rangle+|\bar{2}\rangle\right)\nonumber \\ 
|\psi_{0}\rangle &=U_{3}^{in}|00\rangle =  \frac{1}{\sqrt{3}}\left(|00\rangle+|01\rangle+|10\rangle\right).
\end{align}

In general, the circuit producing the state $|\psi_{0}\rangle$ is hard to find, except when $d$ is a power of 2, as explained above. On the contrary, a circuit implementing the generalized CZ gate is simpler since this gate only introduces a phase in some qudit states and we can reproduce this effect by using controlled-Phase gates, i.e. CPh($\theta)=|00\rangle\langle 00|+|01\rangle\langle 01|+|10\rangle\langle 10|+e^{i\theta}|11\rangle\langle 11|$.

\begin{figure}[t!]
\[
\Qcircuit @C=1.5em @R=.01em @!R 
{
 & \ctrl{3} & \qw & & &
 & \multigate{1}{4} & \qw & & & \qw & \qw & \ctrl{5} & \qw   \\
 &  &  &  &  &  
 &  \ghost{4} & \qw &  &  & \ctrl{4} & \ctrl{3}& \qw & \qw   \\
 &  & &  & & & \qwxo{\hspace{1pt}} & & & & & & &  \\
  & \qwxo{4} & & \equiv & & & \qwxo{\hspace{1pt}} & & = & & & & & \\
 &  &  &  &  &
   & \multigate{1}{4} \qwxo{\hspace{1pt}} & \qw &  &  & 
   \qw & \control \qw & \qw & \qw \\
 & \ctrl{-3} & \qw & & &  &
  \ghost{4} & \qw & & & \gate{S} & \qw & \control \qw & \qw 
}
\]
\caption{Generalized CZ gate for ququarts, $d=4$, performed with four qubits. First gate is a controlled-S gate, which is actually a $CPh$ gate with $\theta=\pi/2$. Last two CZ gates can be implemented in parallel, so the circuit depth is just 2 gates.}
\label{Fig:CZ4}
\end{figure}
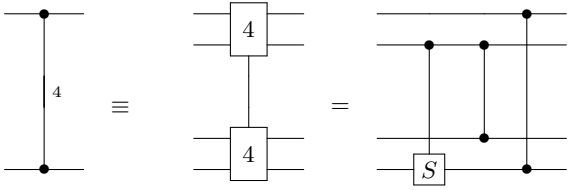

Figure \ref{Fig:CZ3} shows the required circuit to implement generalized CZ gate for qutrits with qubits. Wel need four qubits and four CPh gates to achieve the expected result of this gate. The quantum circuit required to implement the generalized CZ gate for ququarts is shown in Figure \ref{Fig:CZ4}. Only three gates are needed here: two qubit CZ gates and a controlled-S gate, which is a CPh with $\theta=\pi/2$.

At this point, all ingredients to construct the AME states for qubits and to simulate AME states with local dimension $d>2$ has been introduced. Figures \ref{Fig:AME5d} and \ref{Fig:AME6d} can be used to simulate any AME(5,$d$) and AME(6,$d$) state with qubits, providing $U_d^{in}$ and CZ gates. Similarly, Figure \ref{Fig:AME4d} can be used to simulate any AME(4,$d$) state for prime dimension $d\geq3$. Finally, Figure \ref{Fig:AME44_circuit} shows explicitly the circuit and the graph required to obtain the AME(4,4) state.

\begin{figure}[t!]
\[
\Qcircuit @C=0.7em @R=0.001em @!R 
{
 & \ctrl{3} & \qw  & &  &
 & \multigate{1}{3} & \qw & & & \qw  & \ctrl{4} & \ctrl{5} & \qw  & \qw  \\
 &  &  &  &  &  
 &  \ghost{3} & \qw &  &  & \ctrl{4} & \qw & \qw &  \ctrl{3} & \qw  \\
 &  & &  & & & \qwxo{\hspace{1pt}} & & & & & & & & \\
  & \qwxo{3} & & \equiv & & & \qwxo{\hspace{1pt}} & & = & & & & & & \\
 &  &  &  &  &
   & \multigate{1}{3} \qwxo{\hspace{1pt}} & \qw &  &  & 
   \qw & \gate{2\pi/3} & \qw & \gate{-2\pi/3} & \qw \\
 & \ctrl{-3} & \qw & & &  &
  \ghost{3} & \qw & & & \gate{2\pi/3} & \qw & \gate{-2\pi/3} & \qw & \qw 
}
\]
\caption{Generalized CZ gate for qutrits, $d=3$, performed with four qubits. First two CPh gates and last two CPh gates can be implemented in parallel, so the circuit depth is just 2 CPh gates.}
\label{Fig:CZ3}
\end{figure}
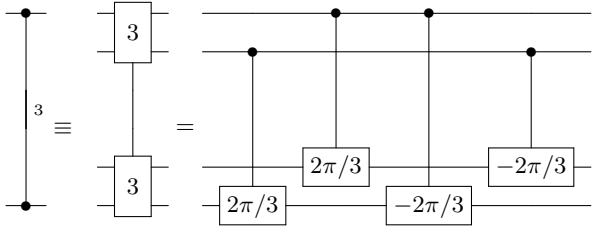

\subsection{AME states circuits of minimal support}

Since AME states of minimal support have connections with error correcting codes, it could be interesting to find the corresponding quantum circuits to generate them. 

\begin{figure}[t!]
\centering
{\[
\Qcircuit @C=1.5em @R=1em @!R {
\lstick{\ket{\bar{0}}} & \qw & \qw & \targ & \qw_{3} & \targ & \targ_{3} & \qw_{3}\\
\lstick{\ket{\bar{0}}} & \qw & \targ & \qw_{3} & \targ & \qw_{3} & \qw & \qw \\
\lstick{\ket{\bar{0}}} & \gate{F_{3}} & \qw & \qw & \ctrl{-1} & \ctrl{-2} & \ctrl{-2} & \qw\\
\lstick{\ket{\bar{0}}} & \gate{F_{3}} & \ctrl{-2} & \ctrl{-3} & \qw & \qw & \qw & \qw
}
\]
}
\caption{Quantum circuit required to generate the state
$|\Omega_{4,3}\rangle$ (4 qutrits) based on the Fourier gate $F_{3}$ and C$_{3}$--adder gate for qutrits. 
}
\label{Fig:AME43}
\end{figure}
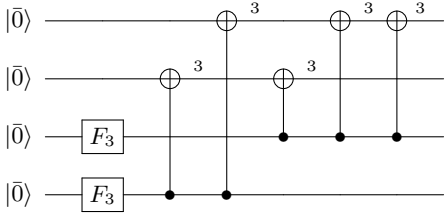

For qutrits, the AME(4,3) state of Eq.\eqref{eq:AME43} has minimal support.
The quantum circuit that generates this state is shown in Figure \ref{Fig:AME43} \cite{GALRZ15}. The quantum gates required to construct this circuit are the Fourier transform gate for qutrits $F_{3}$ and the C$_3$--adder gate
\begin{equation}
\overbar{\mathrm{C}}_{3}|\overbar{i}\rangle|\overbar{j}\rangle=|\overbar{i}\rangle|\overbar{i+j}\rangle,
\end{equation}
which is the generalization of CNOT gate for qutrits. It is represented with the CNOT symbol with the superscript 3, see Figure \ref{Fig:AME43}.

The simulation of the state $F_{3}|\bar{0}\rangle$ by using qubits has been already explained in the previous subsection. 
The construction of the C$_{3}$--adder gate is more cumbersome and we leave the details to the Appendix \ref{app:C3}. The strategy that we use consists in using controlled gates that allow us to perform the sums separately for each control state. If the control qutrit is in the state $|\bar{0}\rangle$ we should apply the identity, so that no gates are needed in this case. If the control qutrit is prepared in the state $|\bar{1}\rangle$, i.e. $|01\rangle$, then we should implement CNOT and Toffoli gates (CCNOT) that take the second qubit as a control qubit, i.e. the second pair of qubits is not affected when the first two are prepared in a different state. Similarly, if the qutrit state is $|\bar{2}\rangle$, i.e. $|10\rangle$, we should search for a sequence of CNOT and CCNOT gates that implement the corresponding sums by using as a control qubit the first qubit.

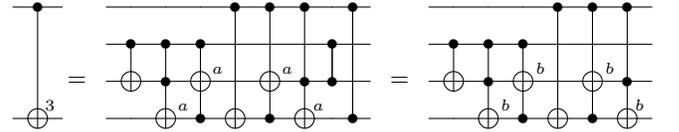
\begin{figure}[h!]
\[
\Qcircuit @C=0.6em @R=.7em @!R 
{
& \ctrl{3} & \qw & & & &  \qw & \qw & \qw & \ctrl{3} & \ctrl{2} & \ctrl{2} & \qw & \ctrl{3} & \qw & & & & &  \qw & \qw & \qw & \ctrl{3} & \ctrl{2} & \ctrl{2} & \qw  \\
 & & & &  & &  \ctrl{1} & \ctrl{1} & \ctrl{1} & \qw & \qw & \qw & \ctrl{1}& \qw & \qw &  & & & &  \ctrl{1} & \ctrl{1} & \ctrl{1} & \qw & \qw & \qw  & \qw \\
 & & & =  & & & \targ & \ctrl{1} & \targ & \qw_{a} & \targ & \ctrl{1}_{a} & \ctrl{-1} & \qw & \qw & & = & & & \targ & \ctrl{1} & \targ & \qw_{b} & \targ & \ctrl{1}_{b} & \qw \\
& \targ & \qw_{3} & & & & \qw & \targ & \ctrl{-1}_{a} & \targ & \ctrl{-1} & \targ & \qw_{a} & \ctrl{-3} & \qw & & & & & \qw & \targ & \ctrl{-1}_{b} & \targ & \ctrl{-1} & \targ & \qw_{b}
}
\]
\caption{C$_{3}$--adder implemented with approximate Toffoli gates of Fig. \ref{Fig:Toffaprox}. The C$_{3}$--adder that uses the CCNOT$_a$ gates needs extra controlled-Z gates to cancel out the minus signs introduced by the approximation.}
\label{Fig:C3adder_approx}
\end{figure}

\begin{figure}[t!]
\[
\Qcircuit @C=0.5em @R=0em @!R {
& \ctrl{1} & \qw & & & & & \ctrl{1} & \qw & & & & &  \qw & \qw & \qw & \ctrl{2} & \qw & \qw & \qw & \qw\\
& \ctrl{1} & \qw & & \simeq & & & \ctrl{1} & \qw & & = &  & & \qw & \ctrl{1} & \qw & \qw & \qw & \ctrl{1} & \qw & \qw &  \\
& \targ & \qw & & & & & \targ & \qw_{a} & & & & & \gate{R_{Y}^{-3\pi/2}} & \ctrl{-1} & \gate{R_{Y}^{3\pi/2}} & \ctrl{-2} & \gate{R_{Y}^{-3\pi/2}} & \ctrl{-1} & \gate{R_{Y}^{3\pi/2}} & \qw
} \]\\
\[
\Qcircuit @C=0.5em @R=0em @!R {
& \ctrl{1} & \qw & & & & & \ctrl{1} & \qw & & & & &  \qw & \qw & \qw & \ctrl{2} & \qw & \qw & \qw & \qw\\
& \ctrl{1} & \qw & & \simeq & \quad  & & \ctrl{1} & \qw & & = & & & \qw & \ctrl{1} & \qw & \qw & \qw & \ctrl{1} & \qw & \qw &  \\
& \targ & \qw & & & & & \targ & \qw_{b} & & & & & \gate{R_{Y}^{\pi/4}} & \targ & \gate{R_{Y}^{\pi/4}} & \targ & \gate{R_{Y}^{-\pi/4}} & \targ & \gate{R_{Y}^{-\pi/4}} & \qw
}
\]
\caption{Approximations of CCNOT gate. They introduce a change of sign in some states, in particular $\mathrm{CCNOT}_{a}|101\rangle =  -|101\rangle$ and $\mathrm{CCNOT}_{b}|100\rangle = -|100\rangle$.}
\label{Fig:Toffaprox}
\end{figure}
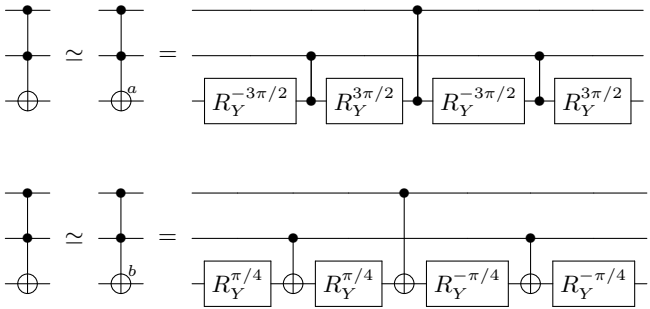

The resulting circuit is depicted in Figure \ref{Fig:C3adder_approx}, where we have used approximate CCNOT gates described in Figure \ref{Fig:Toffaprox}, CCNOT$_{a}$ and CCNOT$_{b}$, instead of usual CCNOT gates in order to reduce significantly the circuit depth \cite{Barenco}. This circuit is divided in two sectors, each one performing the C$_{3}$--adder gate if the controlled qubit is $|\bar{1}\rangle$, the first 3 gates, or $|\bar{2}\rangle$, the last 3 gates. Any of those gates affect the qubit state if the control qutrit is in the $|\bar{0}\rangle$ state.

Clearly, gate C$_{3}$ is the responsible for the growth of circuit depth. However, we can implement the first two adders by using two CNOT gates each one taking advantage that the target qutrit state is $|\bar{0}\rangle$, i.e. qubits are prepared in the state $|00\rangle$.

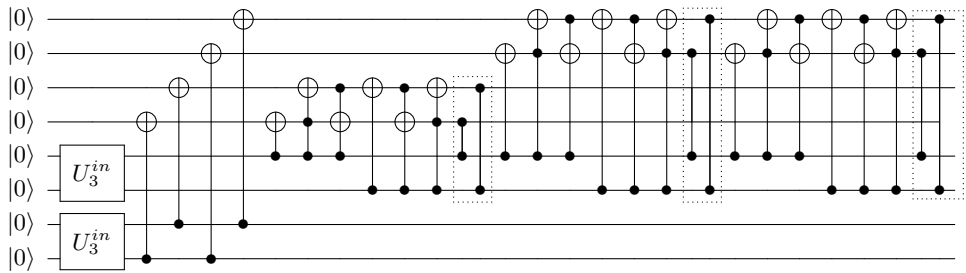
\begin{figure*}[!ht]
\[
\Qcircuit @C=0.5em @R=0.5em @!R {
\lstick{\ket{0}} & \qw & \qw & \qw & \qw & \targ & \qw & \qw & \qw & \qw & \qw & \qw & \qw & \qw & \qw & \targ & \ctrl{1} & \targ & \ctrl{1} & \targ & \qw & \ctrl{5} & \qw & \targ &\ctrl{1} & \targ & \ctrl{1} & \targ & \qw & \ctrl{5}& \qw \\
\lstick{\ket{0}} & \qw & \qw & \qw & \targ & \qw & \qw & \qw & \qw & \qw & \qw & \qw & \qw & \qw & \targ & \ctrl{-1} & \targ & \qw & \targ & \ctrl{-1} & \ctrl{2} & \qw & \targ & \ctrl{-1} & \targ & \qw & \targ & \ctrl{-1} & \ctrl{2} & \qw & \qw \\
\lstick{\ket{0}} & \qw & \qw & \targ & \qw & \qw & \qw & \targ & \ctrl{1} & \targ & \ctrl{1} & \targ & \qw & \ctrl{3} & \qw & \qw & \qw & \qw & \qw & \qw & \qw & \qw & \qw & \qw & \qw & \qw & \qw & \qw & \qw & \qw & \qw \\
\lstick{\ket{0}} & \qw & \targ & \qw & \qw & \qw & \targ & \ctrl{-1} & \targ & \qw & \targ & \ctrl{-1} & \ctrl{1} & \qw & \qw & \qw & \qw & \qw & \qw & \qw & \qw & \qw & \qw & \qw & \qw & \qw & \qw & \qw & \qw & \qw \\
\lstick{\ket{0}} & \multigate{1}{U_{3}^{in}} & \qw & \qw & \qw & \qw & \ctrl{-1} & \ctrl{-1} & \ctrl{-1} &  \qw & \qw & \qw & \ctrl{-1} & \qw & \ctrl{-3} & \ctrl{-3} & \ctrl{-3} & \qw & \qw & \qw & \ctrl{-2} & \qw & \ctrl{-3} & \ctrl{-3} & \ctrl{-3} & \qw & \qw & \qw & \ctrl{-2} & \qw & \qw  \\
\lstick{\ket{0}}  & \ghost{U_{3}^{in}} & \qw & \qw & \qw & \qw  & \qw & \qw & \qw & \ctrl{-3} & \ctrl{-2} & \ctrl{-2} & \qw & \ctrl{-3} & \qw & \qw & \qw & \ctrl{-5} & \ctrl{-4} & \ctrl{-4} & \qw & \ctrl{-5} & \qw & \qw & \qw & \ctrl{-5} & \ctrl{-4} & \ctrl{-4} & \qw & \ctrl{-5} & \qw & \\
\lstick{\ket{0}} & \multigate{1}{U_{3}^{in}} & \qw & \ctrl{-4} & \qw & \ctrl{-6} & \qw & \qw & \qw & \qw & \qw & \qw & \qw & \qw & \qw & \qw & \qw & \qw & \qw & \qw & \qw & \qw & \qw & \qw & \qw & \qw & \qw & \qw & \qw & \qw & \qw \\
\lstick{\ket{0}}  & \ghost{U_{3}^{in}} & \ctrl{-4} & \qw & \ctrl{-6} & \qw & \qw & \qw & \qw & \qw & \qw & \qw & \qw & \qw & \qw & \qw & \qw & \qw & \qw & \qw & \qw & \qw & \qw & \qw & \qw & \qw & \qw & \qw & \qw & \qw & \qw
\gategroup{3}{13}{6}{14}{0.7em}{.}
\gategroup{1}{21}{6}{22}{0.7em}{.}
\gategroup{1}{29}{6}{31}{0.7em}{.}
}
\]
\caption{Circuit for the construction of the AME(4,3) state by using two qubits to represent each qutrit. The controlled-Z gates (framed with dots), are only necessary when we use the approximation of Toffoli gate CCNOT$_a$.}
\label{Fig:AMEqubits}
\end{figure*}

The final circuit required to simulate the state $|\Omega_{4,3}\rangle$ with qubits is shown in Figure \ref{Fig:AMEqubits}, where CZ gates are framed because they are only necessary if we are implementing the CCNOT$_a$ gate. 

\section{Entanglement majorization}\label{sec:major}

Majorization has deep implications in quantum information theory \cite{Nielsen}. In particular, quantum algorithms obey a majorization arrow, which means that majorization could be at the core of their efficiency \cite{LatorreMartin,OLM02}. Following this idea, we wonder whether the above quantum circuits designed to construct AME states obey majorization. If not, it is interesting to asking whether more efficient circuits obeying majorization exist.

Let $\mathbf{a},\mathbf{b}\in \mathbb{R}^{d}$ be vectors having entries ordered in decreasing order, namely $\mathbf{a}^{\downarrow}$ and $\mathbf{b}^{\downarrow}$ with $a_{i+1}^{\downarrow}\geq a_{i}^{\downarrow}$, and similarly for $\mathbf{b}^{\downarrow}$. We say that $\mathbf{a}$ \emph{majorizes} $\mathbf{b}$, i.e. $\mathbf{a}\succ\mathbf{b}$, iff
\begin{eqnarray}
\sum_{i=1}^{k}a_{i}^{\downarrow}\geq\sum_{i=1}^{k}b_{i}^{\downarrow} \quad \mathrm{for} \ k=1,\cdots, d,
\label{eq:majorization}
\end{eqnarray} 
and $\sum_{i=1}^{d}a_{i}=\sum_{i=1}^{d}b_{i}$.

First, we should choose a set of parameters to study if they majorize at each step during the computation, i.e. after the application of each CZ gate. Since all circuits start with a product state and finish with a maximally entangled state in all bipartitions, a natural choice will be the eigenvalues of the reduce density matrices. At some step $s$ during the computation, the circuit has generated a quantum state with density matrix $\rho_{s}$. We then compute the reduce density matrix of every of its bipartitions in two subsytems, $A$ and $B$, i.e. $\rho_{A}^{s}=\mathrm{Tr}_{B}\rho_{s}$, and diagonalize this matrix to obtain its eigenvalues $\mathbf{\lambda^s}=\{\lambda_{i}^{s}\}$. We will establish that this circuit obeys majorization iff $\mathbf{\lambda^s}\succ\mathbf{\lambda^{s+1}}$, i.e.
\begin{equation}
\sum_{i=1}^{k}\left(\lambda_{i}^{\downarrow}\right)^{s} \geq  \sum_{i=1}^{k}\left(\lambda_{i}^{\downarrow}\right)^{s+1} \ \mathrm{for} \ k=1,\cdots ,d^{m}-1 \ \forall A, s,
\end{equation}
where $m=n-\floor{n/2}$ is the number of qudits in $A$ bipartition. We do not consider last summation $k=d^m$ because the eigenvalues of a density matrix are normalized to the unity. Since there are $\left(\begin{array}{c}n\\ \floor{n/2} \end{array}\right)$ bipartitions, this analysis leads to a total number of $\left(\begin{array}{c}n\\ \floor{n/2} \end{array}\right)(d^{m}-1)$ inequalities to fulfill.

\begin{figure}[t!]
\subfloat[]{
\includegraphics[width=7cm, height=4cm]{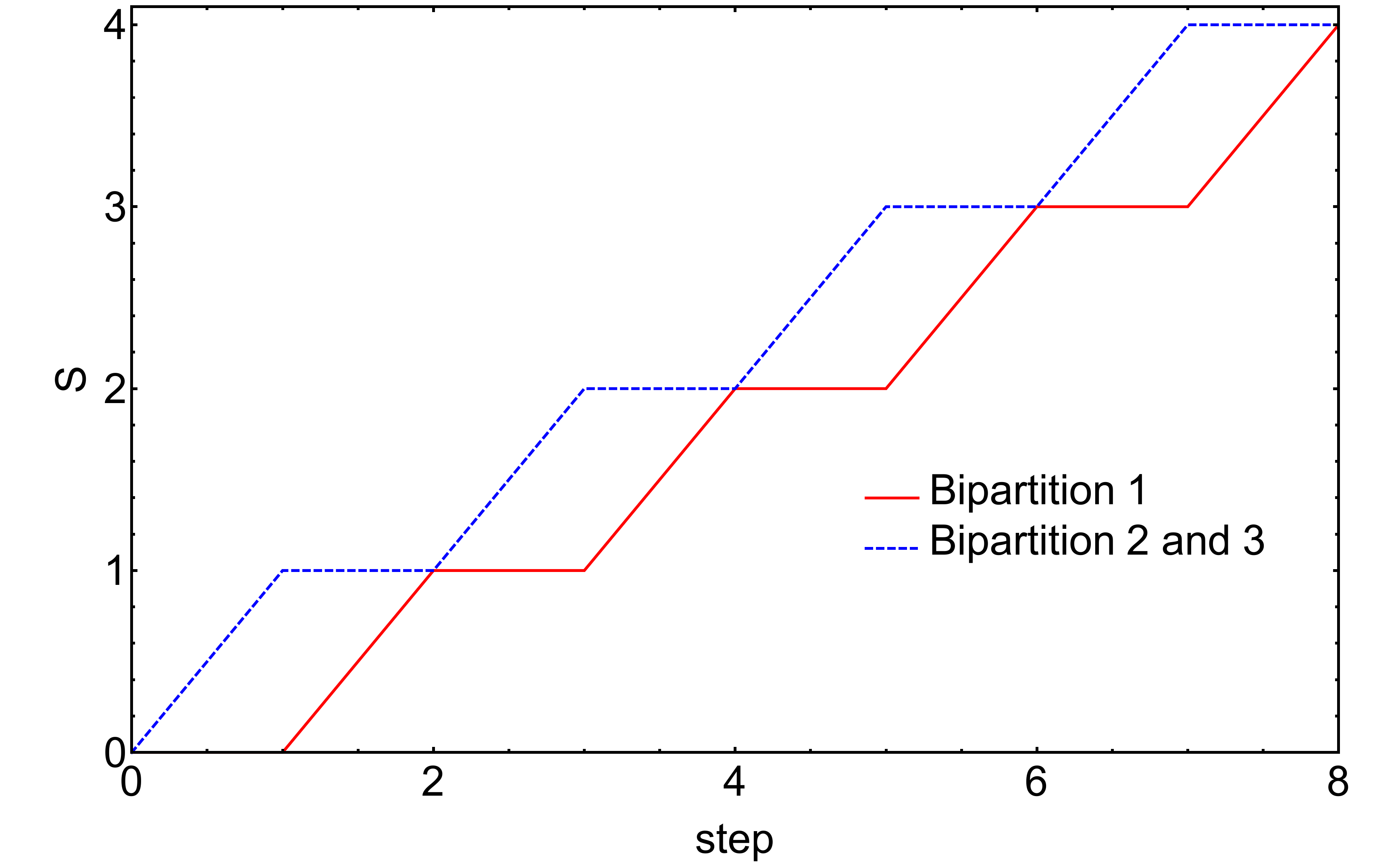}
\label{fig:majS}}\\
\subfloat[]{
\includegraphics[width=6.5cm, height=4cm]{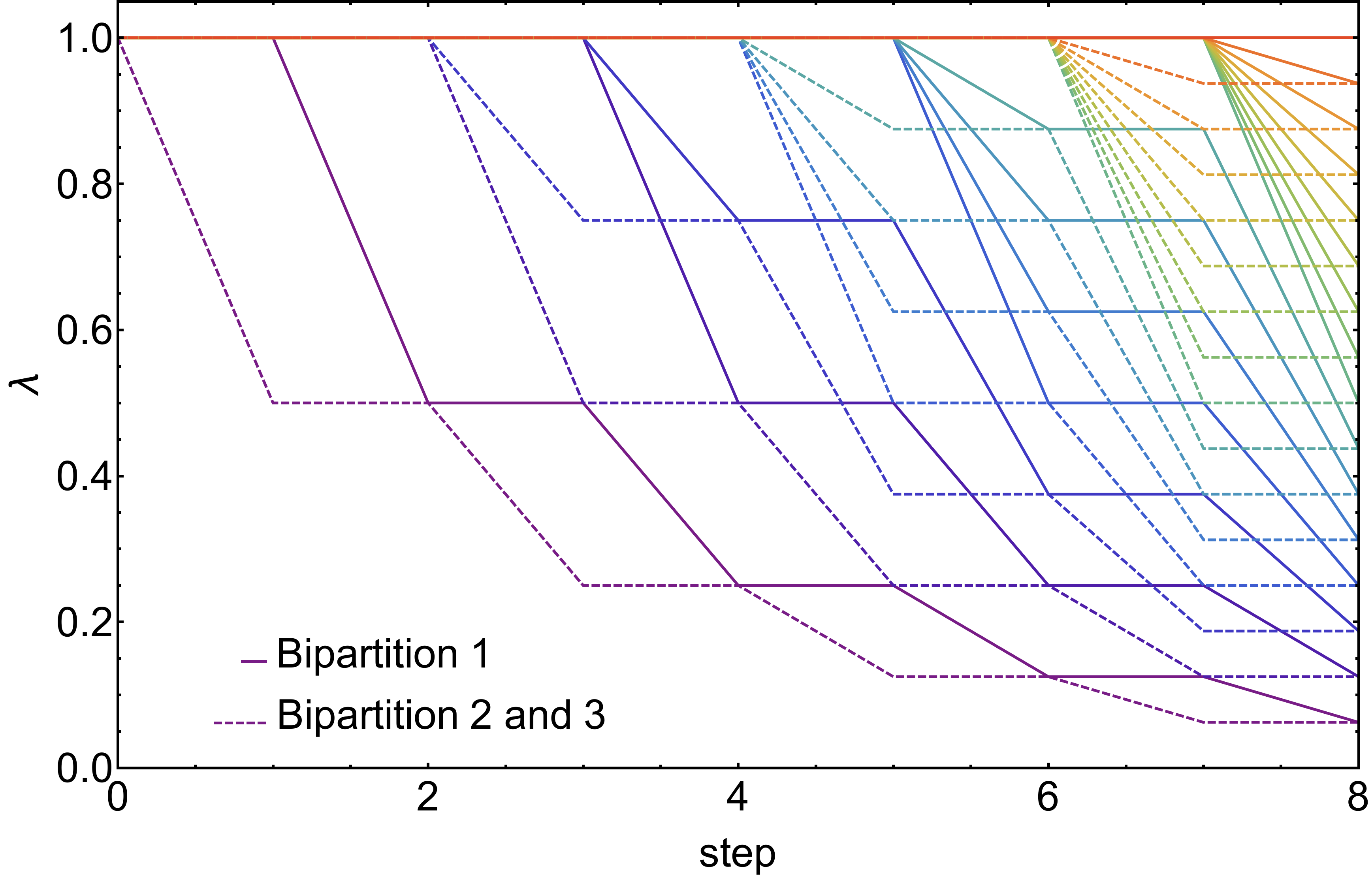}
\label{fig:maj}}
\caption{Majorization in AME(4,4) state circuit of Fig. \ref{Fig:AME44_circuit}. Entropy increases at each step $s$ in all bipartitions until it reach the maximum value $S=2\log_{2}4=4$ (a). Majortization in terms of eigenvalues of the reduce density matrix. At the end of the computation, all eigenvalues are the same, which leads to a density matrix proportional to the identity (b).}
\label{Fig:major}
\end{figure}

We can apply less strict tests by looking at the majorization of other figures of merit to quantify bipartite entanglement, for instance Von Neumann entropy or purity, which in terms of $\lambda_{i}$ are defined as $S=-\sum_{i}\lambda_{i}\log_{d}\lambda_{i}$ and $\gamma=\sum_{i}\lambda_{i}^2$ respectively. Both of these functions are convex in terms of $\lambda_{i}$, so we can apply the Karamata's inequality \cite{Karamata} to prove that
\begin{eqnarray}
\mathbf{\lambda^s}\succ\mathbf{\lambda^{s+1}} &\Rightarrow & S^{s}\leq S^{s+1} \label{eq:majS}\nonumber\\
&\Rightarrow & \gamma^{s}\geq \gamma^{s+1}. \label{eq:majP}
\end{eqnarray}
Thus, we can first do one of these less restrictive tests. If the above inequalities are not fulfilled in every step, then there is no majorization in eigenvalues.

As an example, Figure \ref{Fig:major} shows the majorization of AME(4,4) state of Figure \ref{Fig:AME44_circuit} in terms of entropy and eigenvalues of the reduce density matrix for each bipartition. The circuit majorizes since entropy never decreases and eigenvalues never increase at each step. At the end of the computation, all bipartition have reached the maximum value $S=2\log_{2}4=4$ when all eigenvalues are identical, meaning that reduced density matrices are proportional to the identity, as expected for an AME state.

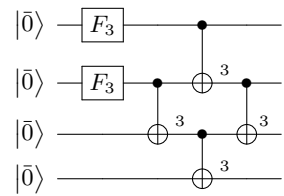
\begin{figure}[t!]
\centering
\[
\Qcircuit @C=1em @R=1em 
{
 \lstick{\ket{\overbar{0}}} & \gate{F_3} & \qw & \ctrl{1} & \qw & \qw \\
 \lstick{\ket{\overbar{0}}} & \gate{F_3} & \ctrl{1} & \targ & \ctrl{1}_{3} & \qw \\
 \lstick{\ket{\overbar{0}}} & \qw & \targ & \ctrl{1}_{3} & \targ & \qw_3 \\
 \lstick{\ket{\overbar{0}}} & \qw & \qw & \targ & \qw_{3} & \qw 
}\]
\caption{Quantum circuit to obtain an AME(4,3) of minimal support. This circuit has been found after applying a majorization test in circuit of Fig. \ref{Fig:AME43}. The number of C$_3$--adder gates and circuit depth have been reduced in one unit, since two of these gates can be applied in parallel. For the qubit simulation, this is a significant gain in terms of circuit complexity.
}
\label{Fig:AME43_optim}
\end{figure}

After analyzing the circuit to construct the state $|\Omega_{4,3}\rangle$ written in Figure \ref{Fig:AME43}, we found that it does not majorize, i.e. when the fourth C$_3$--adder is applied, the entropy of one of the bipartitions decreases before reaching the maximum value after the application of the last C$_3$--adder gate. For this reason, we conclude that this circuit is not optimal, being possible to obtain an AME(4,3) state with minimal support from a smaller number of gates. In particular, we found many equivalent circuits that can obtain this kind of state with only four C$_3$--adder gates. An example is shown in Figure \ref{Fig:AME43_optim}. 
Notice that, in this example, two C$_3$--adders are applied in parallel, which reduces significantly the circuit depth, specially if we want to simulate this AME with qubits.

We found that circuits for AME($n,d$) states majorize up to $n=6$ and $d=4$, with exception of AME(6,2) and AME(6,4). In these two cases, only one bipartition does not majorize, which shows the high optimality of the entanglement power of the circuits proposed. 

One can use this majorization criteria to find optimal entangling circuits based on graph states. For instance, if we are interested in entangle eight parties of our circuit, we can construct a greedy algorithm that finds such a circuit by imposing entanglement majorization. Moreover, we can restrict this algorithm to the given chip architecture, making it suitable for the experimental implementation.

%

\section{Experimental implementation}\label{sec:implementation}

The experimental implementation of an AME state is a highly demanding task for a quantum computer. It requires the consideration of some figure of merit in order to test the quality of preparation state. For qubit AME states of bi-partite and three-partite systems one can consider Mermin Bell inequalities as a figure of merit, as they are maximally violated by these states \cite{Mermin90}. On the other hand, for AME(5,2), AME(6,2) and any qubit graph state in general, there exist Bell inequalities maximally violated by these states \cite{GTHB05}. Besides Bell inequalities, one can also implement a quantum tomography protocol to reconstruct 
the state, being the fidelity of state reconstruction the figure of merit. 
This kind of protocols typically require a quadratic number of measurements outcomes, as a function of the dimension of the Hilbert space \cite{WF89,RBSC04,S06}. However, this number can be reduced to scale linearly with the dimension when \emph{a priori} information is available, e.g. when the state is nearly pure \cite{GCEGXLD15}.

As a first attempt to test the quality of implementation of AME states in quantum computers we considered a very simple test: check whether probability outcomes associated to a measurement in the computational basis are similar to theoretical probabilities. This is not a refined test, as complex phases of entries also play a crucial role. However, a suitable behavior of probabilities along a single projective measurement is a first indication that the state could be successfully prepared.

\begin{figure}
\centering
\includegraphics[width=0.4\textwidth]{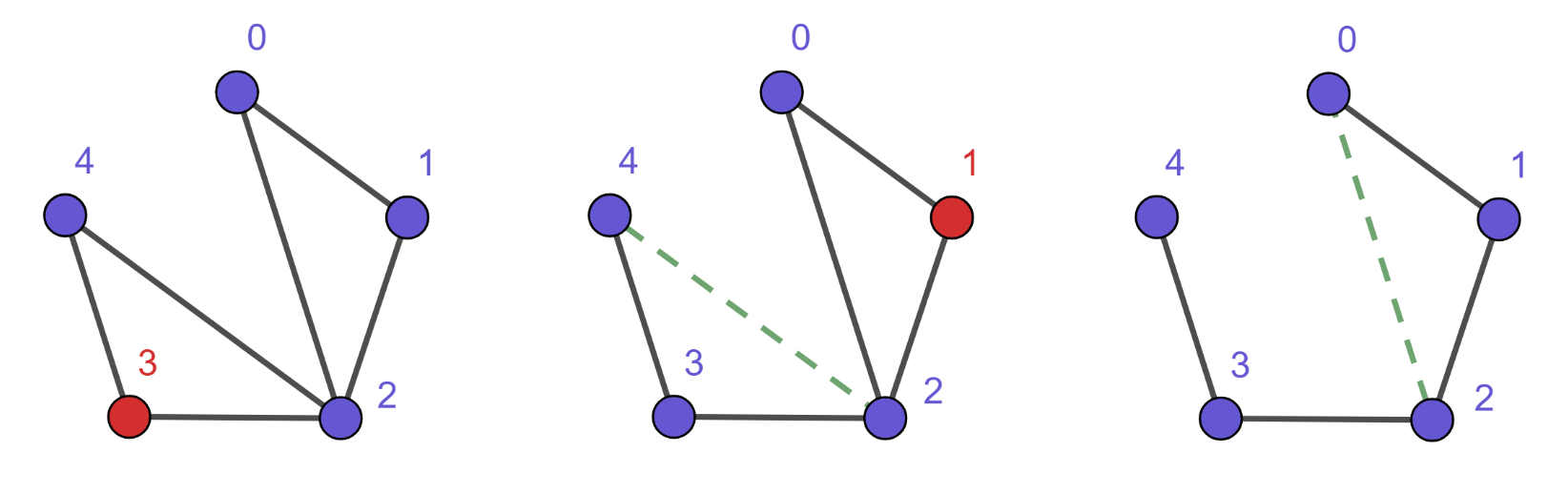}
\caption{Left graph shows the ibmqx4 connectivity. After applying LU operations, one can transform this graph into the linear graph, which belongs to a different graph state class than the one that includes the AME state \cite{HEB04}. The recipe is the following: taking one vertex (in red), one connects all vertices that are connected with the selected one or, in case they are connected, one erase these edges (dashed green lines). This result means that more connections are necessary in order to generate the AME(5,2) graph state in this device.}
\label{Fig:IBM_topo}
\end{figure}

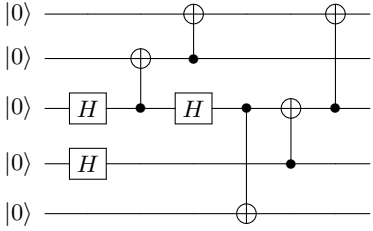
\begin{figure}[t!]
\centering
\[
\Qcircuit @C=1em @R=1em 
{
 \lstick{\ket{0}} & \qw        & \qw       & \targ        & \qw     & \qw  & \targ  & \qw\\
 \lstick{\ket{0}} & \qw        & \targ     & \ctrl{-1}    & \qw      & \qw  & \qw  & \qw\\
 \lstick{\ket{0}} &\gate{H}  & \ctrl{-1} & \gate{H}  & \ctrl{2} & \targ  & \ctrl{-2} & \qw\\
 \lstick{\ket{0}} & \gate{H}  & \qw       & \qw          & \qw   & \ctrl{-1} & \qw & \qw\\
 \lstick{\ket{0}} & \qw         & \qw       & \qw          & \targ    & \qw  & \qw & \qw
}\]
\caption{Circuit to generate an AME(5,2) state on the ibmqx4 quantum computer provided by IBM. We optimized the circuit according to connectivity restriction, in the sense of minimazing the number of entangling gates. The minimal circuit depth achieved is not possible to reproduce when considering the graph AME state.
}
\label{Fig:AME52_IBM}
\end{figure}

We have run two different circuits to generate AME(5,2) state in two quantum computers: the ibmqx4 device from IBM \cite{IBMspec} and the Acorn device from Rigetti Computing \cite{Acorn}. Due to connectivity restriction, it is not possible to implement the simplest quantum circuits predicted by graph states. For instance, the ibmqx4 chip needs from at least one extra CZ gate, as shows Figure \ref{Fig:IBM_topo}. We were able to generated an AME(5,2) state composed by five entangling gates and taking into account the restricted connectivity. The circuit is shown in Figure \ref{Fig:AME52_IBM}. For the Rigetti device, we were not able to find a circuit composed by five entangling gates, so we had to adapt the AME(5,2) graph state to the restricted connectivity by using SWAP gates.

The AME(5,2) state of Figure \ref{Fig:AME52_IBM} is given by
\begin{align}\label{AME52Fig16}
|\mathrm{AME}_{5,2}\rangle=\frac{1}{2\sqrt{2}}&\big(|00000\rangle + |00011\rangle + |01101\rangle + |01110\rangle+ \nonumber\\
&  |10101\rangle + |10110\rangle + |11000\rangle + |11011\rangle\big).
\end{align}
The theoretical probability $P_{ijklm}$ of obtaining each element of the 5-qubit computational basis $|ijklm\rangle$ shown in (\ref{AME52Fig16}) is $1/8 = 0.125$. The results obtained after running the circuit of Figure \ref{Fig:AME52_IBM} in the ibmqx4 device, when considering 8192 shots, are the following:
\begin{equation}
\begin{array}{cc}
P_{00000} = 0.105, &\hspace{0,5cm} P_{00011} = 0.058, \\ P_{01101} = 0.038, & \hspace{0,5cm}P_{01110} = 0.128, \\
P_{10101} = 0.035, &\hspace{0,5cm} P_{10110} = 0.135, \\ P_{11000} = 0.084, &\hspace{0,5cm} P_{11011} = 0.052,
\end{array}
\end{equation}
where $|\psi\rangle$ is the real quantum state generated by the quantum device.
It seems that only three element basis are well reproduced, namely $|10110\rangle$, $|01110\rangle$ and $|00000\rangle$. In addition, two detected probabilities are not related to the AME(5,2) state (\ref{AME52Fig16}), namely $P_{00010} = 0.050$ and $P_{00110}=0.042$.  These imprecise results do not allow us to efficiently implement the adaptative tomographic method presented in Ref. \cite{GCEGXLD15}, as it requires a faithful identification of the highest weights when measuring along the computational basis.

The results with Acorn chip from Rigetti computing were even worst, not allowing us to distinguish results from white noise state preparation. A possible explanation of the failure is related to the large circuit depth due to the consideration of SWAP gates.

The above results illustrate the difficultly to successfully implement AME states on currently existing quantum computers. , the large amount of genuine entanglement required by the states imply a fast decoherence process, reflected even when demanding effectiveness  simple measurement in the computational basis. Our experiment reveals that there are two possible factors involved: \emph{i}) although the quantum circuit of Figure \ref{Fig:AME52_IBM} looks simple, a fast decoherence process occurs due to the high amount of multipartite entanglement required \emph{ii}) the difficulty to successfully implement the challenging state is due to physical limitations of the chip.

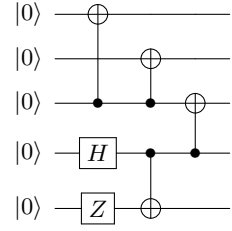
\begin{figure}[t!]
\centering
\[
\Qcircuit @C=1em @R=1em 
{
 \lstick{\ket{0}} & \targ & \qw & \qw & \qw\\
 \lstick{\ket{0}} & \qw & \targ & \qw & \qw\\
 \lstick{\ket{0}} &\ctrl{-2} & \ctrl{-1} & \targ & \qw \\
 \lstick{\ket{0}} & \gate{H} & \ctrl{1} & \ctrl{-1}   & \qw \\
 \lstick{\ket{0}} & \gate{Z} & \targ & \qw & \qw 
}\]
\caption{Quantum circuit required to prepare the 5 qubit GHZ state, restricted to the architecture imposed by the 5-qubit IBM quantum computer ibmqx4. It is worth to mention that the experiment has been implemented in December 2017. Nowadays, the restricted architecture of the computer ibmqx4 has changed.}
\label{Fig:GHZ5}
\end{figure}

Additionally, we implemented the GHZ state in the 5-qubit IBM quantum computer ibmqx4, in order to test violation of the 5-qubit Mermin Bell inequality 
\begin{align}
M_5=&-(a_1a_2a_3a_4a_5) +(a_1a_2a_3a'_4a'_5 + a_1a_2a'_3a_4a'_5  \nonumber\\
&+ a_1a'_2a_3a_4a'_5 +a'_1a_2a_3a_4a'_5 + a_1a_2a'_3a'_4a_5 \nonumber\\
&+ a_1a'_2a_3a'_4a_5 +a'_1a_2a_3a'_4a_5 + a_1a'_2a'_3a_4a_5 \nonumber\\
&+ a'_1a_2a'_3a_4a_5+ a'_1 a'_2 a_3 a_4 a_5 ) \nonumber\\
&-(a_1a'_2a'_3a'_4a'_5 + a'_1a_2a'_3a'_4a'_5 + a'_1a'_2a_3a'_4a'_5\nonumber\\ 
&+a'_1a'_2a'_3a4a'_5 + a'_1a'_2a'_3a'_4a_5) ,
\end{align}
where $a_j$ and $a'_k$ denote two dichotomic observables for five quantum observers \cite{Mermin90}. The theoretical state achieving the maximal violation of the inequality is the GHZ state depicted in Figure \ref{Fig:GHZ5}. This inequality has a classical value $C=4$ and a quantum value $Q=16$. Optimal settings are given by $a_j=\sigma_x$ and $a'_k=\sigma_y$, for $j,k=1,5$. Despite the shortness of the circuit shown in Figure \ref{Fig:GHZ5}, the strong correlations demanded by genuine entanglement imply a fast decoherence process, reflected in a reduction of the strength of violation of the inequality. Nonetheless, the experimentally achieved violation $Q_{exp}=6.90\pm0.01$ is large enough to confirming the genuine non-local nature of the 5-qubit quantum computer ibmqx4.

\section{Discussion and conclusions}\label{sec:conclusions}
 Quantum computing is a challenging field of research in quantum mechanics that could change the way we do computations in the future. The ultimate goal of a quantum computer is to coherently control a relatively large number of qubits in such a way that a multipartite quantum protocol can be successfully implemented, despite the inherent decoherence of quantum information. It is naturally to expect that quantum over classical advantage in computing is directly related to the amount of quantum correlations existing in the involved qubits. It is thus a remarkably important task to understand the behavior of quantum computers when multipartite correlations take extreme values, e.g. when of the system is a genuinelly multipartite maximally entangled state. 
 
In this work, we studied the simplest possible ways to implement genuinely multipartite maximally entangled quantum states, so-called absolutely maximally entangled (AME) states, in order to test the strength of quantum correlations in quantum computers. We explicitly showed a collection of quantum circuits required to implement such states in a some simple scenarios composed by a few qubit systems. For higher dimensional Hilbert spaces, where AME states of qubits do not exist, we considered qudit AME states, where every qudit was artificially generated by considering a group of qubits, see Section \ref{sec:circuits}. For instance, the lack of the AME state for 8 qubit systems can be somehow compensated by considering the AME state of 4 ququarts, where every ququart is composed by two qubits. In this way, pairs of qubits are maximally correlated with three complementary pairs of qubits, thus exhibiting a maximal amount of quantum entanglement in a sense, see Figure \ref{Fig:AME44_circuit}.

One of the main problems when trying to prepare a multipartite quantum circuit over a quantum computer having a restricted architecture is the circuit depth. This is so because some bipartite quantum operations --like CNOT-- are forbidden for some pairs of qubits, as they cannot communicate directly. This physical limitation considerably extends the length of quantum circuit, as typically one has to consider swap operations to complement the lack of communication. In order to deal with this problem, we designed a tool that finds the optimal quantum circuit required to efficiently implement AME states based on entropic majorization of reductions, see Section \ref{sec:major}. As an interesting observation, optimal quantum circuits for AME states typically admit monotonically increasing entropies of reductions, implying that those states can be efficiently generated with our algorithm in a few steps, see Figure \ref{Fig:major}. In other words, our algorithm finds the minimal number of local and non-local quantum gates required to implement those AME states, taking into account the restrictions imposed by the architecture of a real quantum chip.

As a further step, we implemented the GHZ state of 5 qubits over a 5-qubit quantum computer provided by IBM, where we optimized the circuit according to the restrictions imposed by the architecture. The figure of merit to quantify the quality of the state preparation was the violation of the 5-qubit Mermin Bell inequality \cite{Mermin90}, which is maximally violated by the GHZ state. We achieved the experimental non-local value $6.90\pm0.01$, whereas the classical value is $C=4$ and the quantum value is $Q=16$. This result demonstrates the genuine non-local nature of the quantum computer ibmqx4 designed by IBM, which improves a previously achieved quantum value $4.05\pm0.06$ \cite{Dani}. These negative results reflect that the current state of the art of the considered quantum computers is not yet ready to fully exploit the strongest quantum correlations existing in 5 and 6 qubit quantum computers. Nonetheless, we remark that some protocols involving a partial amount of multipartite quantum entanglement have been successfully implemented in quantum computers for a few \cite{D16,M16} and large \cite{B16,B17,Z17} number of qubits.

\section*{Acknowledgements}
We specially thank to Felix Huber and Saverio Pascazio for fruitful comments and discusions. ACL and JIL are supported by Project FIS2015-69167-C2-2-P. DG is supported by Grant FONDECYT Iniciaci\'{o}n number 11180474 and MINEDUC-UA project, code ANT 1855, Chile. The views expressed are those of the authors and do not reflect the official policy or position of IBM or the IBM Quantum Experience team.

\begin{appendix}

\section{C$_3$--adder gate construction}\label{app:C3}

To construct the C$_3$--adder gate with qubits we should find a sequence of gates that perform the following operations:
\begin{align}
\begin{array}{ll}
\mathrm{C}_{3}|00\rangle |00\rangle= |00\rangle |00\rangle, & \mathrm{C}_{3}|01\rangle |00\rangle= |01\rangle |01\rangle, \\
\mathrm{C}_{3}|00\rangle |01\rangle= |00\rangle |01\rangle, & \mathrm{C}_{3}|01\rangle |01\rangle= |01\rangle |10\rangle, \\
\mathrm{C}_{3}|00\rangle |10\rangle= |00\rangle |10\rangle, & \mathrm{C}_{3}|01\rangle |10\rangle= |01\rangle |00\rangle, \\
 & \\
\mathrm{C}_{3}|10\rangle |00\rangle= |10\rangle |10\rangle, & \\
\mathrm{C}_{3}|10\rangle |01\rangle= |10\rangle |00\rangle, & \\
\mathrm{C}_{3}|10\rangle |10\rangle= |10\rangle |01\rangle. &
\end{array}
\label{eq:C3}
\end{align}

As a result, besides from CNOT gates, we will need from CCNOT gates. Three-qubit gates are difficult to implement experimentally, so we should decompose them in terms of one and two-qubit gates. The exact decomposition of CCNOT gate consist on 12 gates of depth. %
%
However, we can use instead an approximate decomposition which differ from the previous for some phase shifts of the quantum states other than zero \cite{Barenco}. In particular, we can use the approximate CCNOT gates shown in Figure \ref{Fig:Toffaprox}. The only changes that those gates introduce respect the exact CCNOT gate are
\begin{align}
\mathrm{CCNOT}_{a}|101\rangle &= -|101\rangle,\nonumber \\
\mathrm{CCNOT}_{b}|100\rangle &= -|100\rangle.
\end{align}
This is translated into the use of controlled-Z gate in the first approximation to obtain the desired result after applying the gate sequence to construct the C$_3$--adder. The sign introduced in the CCNOT$_{b}$ gate is canceled after this sequence, so the circuit remains equal as exact CCNOT gates were used.

We can keep saving more gates. Notice that the firsts two C$_{3}$--adders of the AME circuit of Figure \ref{Fig:AME43} are implemented on qutrits in the state $|\bar{0}\rangle$. Let's write it explicitly. After the Fourier transform on qutrit 1, the circuit applies the C$_{3}$--adder on qutrit 3:
\begin{multline}
(\bar{\mathrm{C}}_{3})_{_{13}}\left[\frac{1}{\sqrt{3}}\left(|\bar{0}\rangle+|\bar{1}\rangle+|\bar{2}\rangle\right)_{1}\otimes|\bar{0}\rangle_{3} \right] \nonumber\\
= \frac{1}{\sqrt{3}}\left(|\bar{0}\bar{0}\rangle+|\bar{1}\bar{1}\rangle+|\bar{2}\bar{2}\rangle\right)_{13},
\end{multline}
where the subindex 13 stands for the qutrits affected from this operation. In qubits form
\begin{multline}
(\mathrm{C}_{3})_{_{13}}\left[\frac{1}{\sqrt{3}}\left(|00\rangle+|01\rangle+|10\rangle\right)_{1}\otimes|00\rangle_{3} \right] \nonumber\\
= \frac{1}{\sqrt{3}}\left(|00\rangle|00\rangle+|01\rangle|01\rangle+|10\rangle|10\rangle\right)_{13}.
\end{multline}
Then, the above operation consists uniquely in two CNOT gates between even and odd qubits. Similarly, the next C$_{3}$--adder acting on qutrit 4 can be implemented in the same way:
\begin{multline}
(\bar{\mathrm{C}}_{3})_{_{14}}\left[\frac{1}{\sqrt{3}}\left(|\bar{0}\bar{0}\rangle+|\bar{1}\bar{1}\rangle+|\bar{2}\bar{2}\rangle\right)_{13}\otimes|\bar{0}\rangle_{4}\right] \nonumber\\
= \frac{1}{\sqrt{3}}\left(|\bar{0}\bar{0}\bar{0}\rangle+|\bar{1}\bar{1}\bar{1}\rangle+|\bar{2}\bar{2}\bar{2}\rangle\right)_{134},
\end{multline}
which in the qubit form becomes
\begin{multline}
(\mathrm{C}_{3})_{_{14}}\left[\frac{1}{\sqrt{3}}\left(|00\rangle|00\rangle+|01\rangle|01\rangle+|10\rangle|10\rangle\right)_{13}\otimes|00\rangle_{4} \right] \nonumber\\
= \frac{1}{\sqrt{3}}\left(|00\rangle|00\rangle|00\rangle+|01\rangle|01\rangle|01\rangle+|10\rangle|10\rangle|10\rangle\right)_{134}.
\end{multline}
Again, the above state can be obtained from the previous using two CNOT gates, between even and odd qubits. This enormous simplification cannot be extended to the other C$_{3}$--adder gates, as all elements of the basis appear once we implement the $F_{3}$ gate on qutrit 2.\\

\end{appendix}

\end{document}